\pdfoutput=1
\documentclass[11pt,a4paper]{article}
\usepackage[T1]{fontenc}
\usepackage[utf8]{inputenc}
\usepackage{lmodern}
\usepackage[margin=1in]{geometry}
\usepackage{amsmath,amssymb,amsthm}
\usepackage{booktabs}
\usepackage{microtype}
\usepackage[colorlinks=true,linkcolor=black,citecolor=black,urlcolor=blue]{hyperref}

\newtheorem{theorem}{Theorem}
\newtheorem{lemma}[theorem]{Lemma}
\newtheorem{corollary}[theorem]{Corollary}
\theoremstyle{remark}
\newtheorem{remark}[theorem]{Remark}

\newcommand{\F}{\mathbb{F}}

\newcommand{\str}[1]{\texttt{#1}}

\title{New linear codes over $\F_7$, $\F_8$ and $\F_9$ from quasi-cyclic codes\\ and Construction~X}
\author{Sebastian Cochinescu\thanks{University of Bucharest, Romania.
ORCID \href{https://orcid.org/0009-0006-1436-4819}{0009-0006-1436-4819}.\newline
E-mail: \texttt{sebastian.cochinescu@drd.unibuc.ro}}}
\date{12 November 2026}

\begin{document}
\maketitle

\begin{abstract}
We present linear codes with parameters $[88,8,64]_7$, $[90,8,65]_7$,
$[98,7,75]_8$, $[97,8,72]_8$, $[44,7,31]_9$, $[97,8,71]_7$, $[85,6,67]_8$
and $[78,8,56]_7$. Each has a minimum distance
one larger than the lower bound listed in Grassl's tables of best known
linear codes; codes with parameters $[89,8,64]_7$, $[96,8,71]_8$,
$[95,8,70]_8$, $[94,8,69]_8$, $[77,8,55]_7$ and $[86,6,68]_8$ follow. Seven of the eight
codes are obtained
from a quasi-cyclic code and a vector that is constant on each block, by
Construction~X with a repetition code. The remaining one, the
$[90,8,65]_7$ code, is obtained from the first by appending two coordinates. The parameters
and all weight counts were obtained by exhaustive enumeration, and a short
program that repeats these computations is included.
\end{abstract}


\section{Introduction}

An $[n,k,d]_q$ code is a $k$-dimensional subspace of $\F_q^n$ in which every
nonzero word has Hamming weight at least $d$, with equality for some word.
For fixed $q$, $n$ and $k$, the largest possible $d$ is known only in part.
The tables maintained by Grassl~\cite{codetables} list, for $q\le 9$, the
best lower and upper bounds known; for $q=7$ they cover $n\le 100$ and for
$q=8$ and $q=9$ they cover $n\le 130$.

Many entries of these tables come from computer searches among constacyclic,
quasi-cyclic and quasi-twisted codes
\cite{chenqt,AydinSiapRay2001,Chen2015,AydinConnollyGrassl2017,Akre2023,Dastbasteh2025},
often followed by standard constructions such as puncturing, shortening,
Construction~X \cite{SloaneReddyChen1972} or the
addition of coordinates~\cite{Kohnert2009}. Short reports of individual new codes are a
common form of publication in this area, for example~\cite{BrownKasprzyk2013}.
This note is of that kind.

\begin{theorem}\label{thm:main}
There exist linear codes with parameters
\[
  [88,8,64]_7,\qquad [90,8,65]_7,\qquad [98,7,75]_8,\qquad [97,8,72]_8,
  \qquad [44,7,31]_9,
\]
\[
  [97,8,71]_7,\qquad [85,6,67]_8,\qquad [78,8,56]_7 .
\]
\end{theorem}

\begin{corollary}\label{cor:89}
There exist linear codes with parameters $[89,8,64]_7$, $[96,8,71]_8$,
$[95,8,70]_8$, $[94,8,69]_8$, $[77,8,55]_7$ and $[86,6,68]_8$.
\end{corollary}

Table~\ref{tab:codes} compares these parameters with the
bounds listed in~\cite{codetables} on 10 November 2026 and, for the last
three codes of Theorem~\ref{thm:main} and for $[77,8,55]_7$ and
$[86,6,68]_8$, on 11 November 2026. In each case the
lower bound increases by one. Codes with the previous parameters
$[88,8,63]_7$ and $[98,7,74]_8$ were given by Daskalov and
Gulliver~\cite{DaskalovGulliver2001}. For $n=97$, $k=8$, $q=8$ the table
lists a $[99,8,73]_8$ code, credited there to T.~Maruta, with two
coordinates deleted; for $n=44$, $k=7$, $q=9$ it lists a $[45,7,31]_9$ code
with one coordinate deleted (see Remark~\ref{rem:E}). For $n=97$, $k=8$,
$q=7$ it lists a quasi-cyclic $[96,8,70]_7$ code with one coordinate
appended; for $n=85$, $k=6$, $q=8$ a $[89,6,70]_8$ code with four
coordinates deleted; and for $n=78$, $k=8$, $q=7$ a quasi-cyclic
$[80,8,57]_7$ code with two coordinates deleted (see
Remarks~\ref{rem:F}, \ref{rem:Gprior} and~\ref{rem:H}). Here ``new'' means that the
parameters exceed these tables on that date. For $[98,7,75]_8$ and
$[44,7,31]_9$, Remarks~\ref{rem:hristov} and~\ref{rem:yu} give a second,
inequivalent code, obtained from a code in the literature by one of the
constructions used here. None of the fourteen codes is known to be optimal.
The first version of this note, of 10 November 2026, contained the first
five codes of Theorem~\ref{thm:main} and the first four of
Corollary~\ref{cor:89}; the other five codes were found after it had been
submitted, two of them later on the same day.

\begin{table}[ht]
\centering
\begin{tabular}{lcccl}
\toprule
Code & \multicolumn{2}{c}{Bounds in \cite{codetables}} & Words of & Construction \\
 & lower & upper & minimum weight & \\
\midrule
$[88,8,64]_7$ & 63 & 69 & 3\,828 & Section~\ref{sec:A} \\
$[90,8,65]_7$ & 64 & 71 & 1\,002 & Section~\ref{sec:B} \\
$[98,7,75]_8$ & 74 & 80 & 4\,256 & Section~\ref{sec:C} \\
$[97,8,72]_8$ & 71 & 79 & 3\,990 & Section~\ref{sec:D} \\
$[44,7,31]_9$ & 30 & 34 & 2\,352 & Section~\ref{sec:E} \\
$[97,8,71]_7$ & 70 & 77 & 3\,876 & Section~\ref{sec:F} \\
$[85,6,67]_8$ & 66 & 71 & 4\,998 & Section~\ref{sec:G} \\
$[78,8,56]_7$ & 55 & 61 & 4\,560 & Section~\ref{sec:H} \\
\midrule
$[89,8,64]_7$ & 63 & 70 & & Corollary~\ref{cor:89} \\
$[96,8,71]_8$ & 70 & 78 & & Corollary~\ref{cor:89} \\
$[95,8,70]_8$ & 69 & 77 & & Corollary~\ref{cor:89} \\
$[94,8,69]_8$ & 68 & 76 & & Corollary~\ref{cor:89} \\
$[77,8,55]_7$ & 54 & 60 & & Corollary~\ref{cor:89} \\
$[86,6,68]_8$ & 67 & 72 & & Corollary~\ref{cor:89} \\
\bottomrule
\end{tabular}
\caption{The new codes and the previous bounds on the minimum distance.}
\label{tab:codes}
\end{table}

The proofs are computer assisted. Each code is given by a few polynomials,
and each statement about weights was obtained by listing all $q^k$ codewords
($7^8 = 5\,764\,801$, $8^6 = 262\,144$, $8^7 = 2\,097\,152$,
$8^8 = 16\,777\,216$ and $9^7 = 4\,782\,969$). Section~\ref{sec:verification}
describes the independent checks and where the generator matrices are
deposited. Section~\ref{sec:search} states how the codes were found,
including the use of AI tools.

\section{Notation and one lemma}\label{sec:notation}

Words are row vectors. For $a\in\F_q$ and an integer $m$, $a^{m}$ stands
for the letter $a$ repeated $m$ times, and $(u\mid v)$ is the concatenation
of $u$ and $v$.

\paragraph{Quasi-cyclic codes.}
Fix $m$ and identify a polynomial $p(x)=p_0+p_1x+\dots+p_{m-1}x^{m-1}$ with the
vector $(p_0,\dots,p_{m-1})$. For polynomials $p_1,\dots,p_\ell$ of degree less
than $m$ and an integer $k\le m$, let
\[
  Q_k(p_1,\dots,p_\ell)=\bigl\{\,(a p_1 \mid a p_2 \mid \dots \mid a p_\ell)
  \bmod (x^m-1) \;:\; a\in\F_q[x],\ \deg a<k \,\bigr\}\subseteq\F_q^{\ell m}.
\]
It is spanned by the $k$ words
$(x^i p_1\mid\dots\mid x^i p_\ell) \bmod (x^m-1)$, $i=0,\dots,k-1$, that is,
by the first $k$ rows of $\ell$ circulant matrices placed side by side. Let
$h$ be a divisor of $x^m-1$ of degree $k$. If $h\,p_j\equiv 0 \pmod{x^m-1}$
for every $j$, then $Q_k(p_1,\dots,p_\ell)$ is invariant under the
simultaneous cyclic shift of all $\ell$ blocks. Such a code is a
one-generator quasi-cyclic code of index $\ell$, and if its dimension is $k$
then $h$ is its check polynomial~\cite{AydinSiapRay2001}. The quasi-cyclic
codes below have dimension $k$; the ancillary program computes their ranks.

Polynomials are written below as strings of coefficients, lowest degree
first; zero coefficients above the leading one are omitted, so a string can
be shorter than $m$. Over $\F_7$ the digits are the integers modulo $7$. Over
$\F_8=\F_2[w]/(w^3+w+1)$ and over $\F_9=\F_3[w]/(w^2+2w+2)$ the digit \str{0}
is zero and the digit $e\ge 1$ is $w^{e-1}$; in both fields $w$ is a
primitive element. So over $\F_8$ the digits \str{1}, \str{2}, \dots, \str{7}
stand for $1,w,w^2,\dots,w^6$, and over $\F_9$ the digits \str{1}, \dots,
\str{8} stand for $1,w,\dots,w^7$, with $w^4=2$.

\begin{lemma}[Construction X with a repetition code]\label{lem:X}
Let $C_0$ be an $[n,k,d_0]_q$ code, let $v\in\F_q^n\setminus C_0$, and let
$e$ be the smallest weight of a word of $C_0+\langle v\rangle$ that is not in
$C_0$. For every integer $s\ge 0$,
\[
  C=\bigl\{\,(c+tv \mid t^{s}) \;:\; c\in C_0,\ t\in\F_q \,\bigr\}
\]
is an $[n+s,\,k+1,\,\min(d_0,\,e+s)]_q$ code.
\end{lemma}

\begin{proof}
The dimension is $k+1$ because $v\notin C_0$. A word with $t=0$ is a word of
$C_0$ followed by zeros, so these words have minimum weight $d_0$. A word
with $t\ne 0$ consists of a word of $C_0+\langle v\rangle$ outside $C_0$
followed by $s$ nonzero letters, and every such word occurs, so these words
have minimum weight $e+s$.
\end{proof}

For $s\ge 1$ this is Construction~X~\cite{SloaneReddyChen1972}, stated there
for binary codes, applied over $\F_q$ to the pair
$C_0\subset C_0+\langle v\rangle$ with the $[s,1,s]$ repetition code as
auxiliary code. When $e<d_0$, the choice $s=d_0-e$ gives minimum distance
$d_0$ with the fewest added coordinates.

\section{The codes}

\subsection{A \texorpdfstring{$[88,8,64]_7$}{[88,8,64]\_7} code}\label{sec:A}

Let $q=7$, $m=29$ and
\begin{align*}
p_1 &= \str{511123121120153444146565541},\\
p_2 &= \str{2234233360156340230562131121},\\
p_3 &= \str{2061452321263301322322543141}.
\end{align*}
Let $C_1=Q_8(p_1,p_2,p_3)\subseteq\F_7^{87}$, and let $A$ be the code obtained
by appending to the word of each $a(x)$ the letter
$a(1)=a_0+a_1+\dots+a_7$. A generator matrix of $A$ consists of the eight
rows $(x^ip_1\mid x^ip_2\mid x^ip_3)\bmod(x^{29}-1)$, $i=0,\dots,7$, each
followed by the letter $1$.

Let $C_0\subset C_1$ be the subcode of the words with $a(1)=0$, of dimension
$7$. Enumeration of all words of $C_1$ gives:
\begin{itemize}
\item $C_0$ has minimum distance $64$, with 870 words of weight $64$;
\item the smallest weight of a word of $C_1$ outside $C_0$ is $63$, with
      2\,958 such words.
\end{itemize}
The word of $a(x)$ in $A$ is $(c+tv\mid t)$, where $t=a(1)$, $v$ is the word
of the constant polynomial $1$ and $c\in C_0$. By Lemma~\ref{lem:X} with
$d_0=64$, $e=63$ and $s=1$, $A$ is an $[88,8,64]_7$ code. It has
$870+2\,958=3\,828$ words of weight $64$.

\begin{remark}\label{rem:A}
The code $C_1$ is a quasi-cyclic $[87,8,63]_7$ code of index $3$ with check
polynomial $h=(x-1)h_7$, where
$h_7=x^7+6x^6+x^5+2x^4+2x^2+5x+6$ is irreducible. The word of $a=4h_7$ is
$j=(1^{29}\mid 4^{29}\mid 3^{29})$ and has $a(1)=1$. So
$A=\{(c+tj\mid t)\}$ with $j$ constant on each block, and $C_0$ is a
quasi-cyclic $[87,7,64]_7$ code with check polynomial $h_7$.
\end{remark}

\begin{remark}\label{rem:canli}
Chen's database of quasi-twisted codes~\cite{chenqt} lists a quasi-cyclic
$[87,8,63]_7$ code of index $3$ with $m=29$, credited there to A.~Canli
(private communication). Its minimum distance already exceeds the lower bound
$62$ listed in~\cite{codetables} for $n=87$, $k=8$, so the parameters
$[87,8,63]_7$ are not new and are not claimed here. That code has the same
check polynomial $h$ as $C_1$ but is a different code: it has 3\,306 words of
weight $63$, whereas $C_1$ has 2\,958. In the notation above, its three
polynomials in~\cite{chenqt} are
\begin{align*}
&\str{6124013321562404416221},\\
&\str{05224152051103534655412614222},\\
&\str{04515006116330636303655505551}.
\end{align*}
\end{remark}

\subsection{A \texorpdfstring{$[90,8,65]_7$}{[90,8,65]\_7} code}\label{sec:B}

Let $B$ be the code generated by the eight rows of the generator matrix of
$A$ in Section~\ref{sec:A}, with two letters appended to each row: to the
row of $x^i$, $i=0,\dots,7$, the entries in position $i+1$ of
\begin{align*}
c_1&=(1,1,5,5,6,1,2,6),\\
c_2&=(0,1,4,3,1,6,3,3).
\end{align*}
The code $A$ has 3\,828 words of weight $64$ and 9\,222 words of weight
$65$. Enumeration gives:
\begin{itemize}
\item every word of weight $64$ of $A$ has a nonzero appended letter
      (822 words have exactly one, 3\,006 have two);
\item 180 words of weight $65$ of $A$ have two zero appended letters.
\end{itemize}
Appending coordinates does not lower any weight, so every nonzero word of
$B$ has weight at least $65$. Hence $B$ is a $[90,8,65]_7$ code, with
$822+180=1\,002$ words of weight $65$. This is a
$(2,1)$-extension in the sense of Kohnert~\cite{Kohnert2009}.

\subsection{A \texorpdfstring{$[98,7,75]_8$}{[98,7,75]\_8} code}\label{sec:C}

Let $q=8$, $\F_8=\F_2[w]/(w^3+w+1)$, $m=19$ and, in the digit notation of
Section~\ref{sec:notation},
\begin{align*}
g_1 &= \str{14012366321041},\\
g_2 &= \str{6755573351203563251},\\
g_3 &= \str{5706506212216534441},\\
g_4 &= \str{1740035466350573513},\\
g_5 &= \str{1010542215603411613}.
\end{align*}
Let $C_0=Q_6(g_1,\dots,g_5)\subseteq\F_8^{95}$. It is a quasi-cyclic code of
index $5$ whose check polynomial
$h=x^6+w^3x^5+w^6x^4+w^6x^3+w^6x^2+w^3x+1$ is irreducible. Let
\[
  j=(1^{19}\mid (w^6)^{19}\mid (w^6)^{19}\mid (w^5)^{19}\mid 1^{19}).
\]
Enumeration gives:
\begin{itemize}
\item $C_0$ is a $[95,6,75]_8$ code with 2\,394 words of weight $75$
      (the lower bound in~\cite{codetables} for $n=95$, $k=6$ is also $75$;
      see Remark~\ref{rem:hristov});
\item the smallest weight of a word of $C_0+\langle j\rangle$ outside $C_0$
      is $72$, with 1\,862 such words.
\end{itemize}
By Lemma~\ref{lem:X} with $d_0=75$, $e=72$ and $s=3$, the code
\[
  C=\{\,(c+tj\mid t,t,t) \;:\; c\in C_0,\ t\in\F_8\,\}
\]
is a $[98,7,75]_8$ code. It has $2\,394+1\,862=4\,256$ words of
weight $75$. A generator matrix consists of the six rows of $C_0$, each
followed by $0,0,0$, and the row $(j\mid 1,1,1)$.

\begin{remark}\label{rem:hristov}
One-generator quasi-cyclic codes of index $5$ with $m=19$ and the parameters
$[95,6,75]_8$ and $[95,7,72]_8$ were found by Hristov~\cite{Hristov2008}, and
the $[95,6,75]_8$ code listed in~\cite{codetables} is his. In the notation
above, its five polynomials in~\cite{codetables} are
\begin{align*}
&\str{1000001401236632104},\quad \str{1551005444536127665},\quad
\str{0010007635274061617},\\
&\str{3565107577555372612},\quad \str{5413101314516655661}.
\end{align*}
It has the same check polynomial $h$ as $C_0$ but is a different code: it
has 2\,660 words of weight $75$, whereas $C_0$ has 2\,394. The recipe of
Lemma~\ref{lem:X} applies to it as well. Up to scalar multiples there are
$(8^5-1)/7=4\,681$ nonzero vectors that are constant on each block. For
Hristov's code exactly one of them has $e=72$, namely
$(1^{19}\mid (w^5)^{19}\mid (w^5)^{19}\mid (w^6)^{19}\mid (w)^{19})$, and none
has $e\ge 73$. For this vector there are 1\,729 words of weight $72$
outside Hristov's code, and with $s=3$ it gives another $[98,7,75]_8$ code,
which has $2\,660+1\,729=4\,389$ words of weight $75$ and so is not
equivalent to $C$. For $C_0$ exactly two of them have $e=72$, the vector $j$ above and
$(1^{19}\mid (w^4)^{19}\mid (w^2)^{19}\mid (w)^{19}\mid (w)^{19})$, and none has
$e\ge 73$. We noticed the code obtained from Hristov's code only after $C$
had been found from $C_0$, which came from our own search.

The $[95,7,72]_8$ code printed in~\cite{Hristov2008}, which is extended
there to a $[97,7,74]_8$ code by two columns, has in the notation above the
polynomials \str{1224172714221}, \str{537376316524475461},
\str{427137751246236361}, \str{464636266201737621} and
\str{12063126362136021}. It has 266 words of weight $72$.
It therefore differs from $C_0+\langle j\rangle$, which has 1\,862, and from
the code $C_0$ of Section~\ref{sec:D}, which has 532. No vector that is
constant on each block gives $e\ge 70$ when added to it, so the
construction of Section~\ref{sec:D} does not give a $[97,8,72]_8$ code from
it.
\end{remark}

\subsection{A \texorpdfstring{$[97,8,72]_8$}{[97,8,72]\_8} code}\label{sec:D}

Let $q=8$, $m=19$ and
\begin{align*}
g_1 &= \str{2271327647420761},\\
g_2 &= \str{533206036137344161},\\
g_3 &= \str{147745324463365031},\\
g_4 &= \str{504257355577332601},\\
g_5 &= \str{5637737133476425551}.
\end{align*}
Let $C_0=Q_7(g_1,\dots,g_5)\subseteq\F_8^{95}$. It is a quasi-cyclic code of
index $5$ with check polynomial $h=(x+1)h_6$, where
$h_6=x^6+w^5x^5+w^3x^4+w^3x^3+w^3x^2+w^5x+1$ is irreducible. Let
\[
  j=(1^{19}\mid 0^{19}\mid (w^5)^{19}\mid 1^{19}\mid (w^2)^{19}).
\]
Enumeration gives:
\begin{itemize}
\item $C_0$ is a $[95,7,72]_8$ code with 532 words of weight $72$
      (the lower bound in~\cite{codetables} for $n=95$, $k=7$ is also $72$);
\item the smallest weight of a word of $C_0+\langle j\rangle$ outside $C_0$
      is $70$, with 3\,458 such words.
\end{itemize}
By Lemma~\ref{lem:X} with $d_0=72$, $e=70$ and $s=2$, the code
\[
  D=\{\,(c+tj\mid t,t) \;:\; c\in C_0,\ t\in\F_8\,\}
\]
is a $[97,8,72]_8$ code. It has $532+3\,458=3\,990$ words of weight $72$. A
generator matrix consists of the seven rows of $C_0$, each followed by
$0,0$, and the row $(j\mid 1,1)$.

\begin{remark}\label{rem:D}
Since $x+1$ divides $h$, the code $C_0$ already contains a word that is
constant on each block: the word of $a=h_6$ is
$((w^5)^{19}\mid (w^6)^{19}\mid (w^4)^{19}\mid (w^4)^{19}\mid (w^4)^{19})$.
So $C_0+\langle j\rangle$ is the sum of a quasi-cyclic code with the
irreducible check polynomial $h_6$ and a two-dimensional space of words
that are constant on each block. The polynomial $h_6$ is not the check
polynomial of Section~\ref{sec:C}.
\end{remark}

\subsection{A \texorpdfstring{$[44,7,31]_9$}{[44,7,31]\_9} code}\label{sec:E}

Let $q=9$, $\F_9=\F_3[w]/(w^2+2w+2)$, $m=14$ and
\begin{align*}
g_1 &= \str{3661546483071},\\
g_2 &= \str{2724866525051},\\
g_3 &= \str{2726641550711}.
\end{align*}
Let $C_0=Q_6(g_1,g_2,g_3)\subseteq\F_9^{42}$. It is a quasi-cyclic code of
index $3$ whose check polynomial
\[
  h=x^6+w^6x^5+w^2x^3+w^6x+w^4=(x^3+wx^2+w^7x+w^4)(x^3+w^7x^2+w^5x+1)
\]
is the product of two irreducible polynomials of degree $3$. Let
\[
  j=(1^{14}\mid (w^7)^{14}\mid (w^6)^{14}).
\]
Enumeration gives:
\begin{itemize}
\item $C_0$ is a $[42,6,31]_9$ code with 1\,456 words of weight $31$;
\item the smallest weight of a word of $C_0+\langle j\rangle$ outside $C_0$
      is $29$, with 896 such words.
\end{itemize}
By Lemma~\ref{lem:X} with $d_0=31$, $e=29$ and $s=2$, the code
\[
  E=\{\,(c+tj\mid t,t) \;:\; c\in C_0,\ t\in\F_9\,\}
\]
is a $[44,7,31]_9$ code. It has $1\,456+896=2\,352$ words of weight $31$. A
generator matrix consists of the six rows of $C_0$, each followed by $0,0$,
and the row $(j\mid 1,1)$.

\begin{remark}\label{rem:E}
The codes $C_0$ and $C_0+\langle j\rangle$ have the parameters
$[42,6,31]_9$ and $[42,7,29]_9$, which equal the lower bounds listed
in~\cite{codetables} for these lengths and dimensions. The code that the
table lists for $n=44$, $k=7$ is of the same kind as $E$: Construction~X
applied to quasi-cyclic $[42,6,31]_9$ and $[42,7,28]_9$ codes of index $3$
with the repetition code of length $3$ gives a $[45,7,31]_9$ code, and
deleting one coordinate gives $[44,7,30]_9$. Here the larger code has
minimum distance $29$, so a repetition code of length $2$ suffices. The
parameters $[45,7,31]_9$ are those of the table and are not claimed here.
\end{remark}

\begin{remark}\label{rem:yu}
Yu and Zhu~\cite{YuZhu2025} give a $[42,7,30]_9$ code, one above the value
$29$ listed in~\cite{codetables}; its generator matrix is in the public
file~\cite{YuRepo}, where the entries are powers of a root $w$ of
$x^2+2x+2$ and $2$ stands for $w^4$, as here. Deleting coordinates, shortening and adding zero
coordinates give from it no $[44,7,31]_9$ code. While looking for earlier
codes we found that appending the columns
$(1,w^3,0,w^2,w,w^5,2)$ and $(1,0,0,2,w^7,w^7,w^7)$ to the seven rows of
that matrix, in the order of the file, gives another $[44,7,31]_9$ code. It
has 1\,360 words of weight $31$, so it is not equivalent to $E$, which has
2\,352. The code $E$ does not use the code of~\cite{YuZhu2025}.
\end{remark}

\subsection{A \texorpdfstring{$[97,8,71]_7$}{[97,8,71]\_7} code}\label{sec:F}

Let $q=7$, $m=19$ and
\begin{align*}
g_1 &= \str{33051250431201511},\\
g_2 &= \str{213131025241354511},\\
g_3 &= \str{344335265466432031},\\
g_4 &= \str{453353336156163241},\\
g_5 &= \str{126346662654142161}.
\end{align*}
Let $C_0=Q_7(g_1,\dots,g_5)\subseteq\F_7^{95}$. It is a quasi-cyclic code of
index $5$ with check polynomial
\[
  h=(x-1)(x^3+4x^2+x+6)(x^3+3x^2+3x+6),
\]
the product of $x-1$ and two irreducible polynomials of degree $3$. Let
\[
  j=(0^{19}\mid 1^{19}\mid 1^{19}\mid 6^{19}\mid 6^{19}).
\]
The vector $j$ is not in $C_0$. Enumeration gives:
\begin{itemize}
\item $C_0$ is a $[95,7,71]_7$ code with 1\,368 words of weight $71$
      (the lower bound in~\cite{codetables} for $n=95$, $k=7$ is also $71$);
\item the smallest weight of a word of $C_0+\langle j\rangle$ outside $C_0$
      is $69$, with 2\,508 such words.
\end{itemize}
By Lemma~\ref{lem:X} with $d_0=71$, $e=69$ and $s=2$, the code
\[
  F=\{\,(c+tj\mid t,t) \;:\; c\in C_0,\ t\in\F_7\,\}
\]
is a $[97,8,71]_7$ code. It has $1\,368+2\,508=3\,876$ words of weight $71$. A
generator matrix consists of the seven rows of $C_0$, each followed by
$0,0$, and the row $(j\mid 1,1)$.

\begin{remark}\label{rem:F}
The code listed in~\cite{codetables} for $n=97$, $k=8$ is a quasi-cyclic
$[96,8,70]_7$ code of index $4$ with $m=24$, with one coordinate appended.
That $[96,8,70]_7$ code has 3\,528 words of weight $70$. Up to scalar
multiples there are $(7^8-1)/6=960\,800$ nonzero columns that can be
appended to its generator matrix, and for each of them some word of weight
$70$ gets the letter $0$. So no extension of that code by one coordinate has
minimum distance $71$. Quasi-cyclic codes with the parameters $[96,8,70]_7$
and $[100,8,73]_7$ are given by Hristov~\cite{Hristov2009}. His
$[100,8,73]_7$ code, of index $4$ with $m=25$, has 2\,100 words of weight
$73$, and every three coordinates are all nonzero in one of these words. So
deleting three coordinates of that code always gives minimum distance $70$.
\end{remark}

\subsection{An \texorpdfstring{$[85,6,67]_8$}{[85,6,67]\_8} code}\label{sec:G}

Let $q=8$, $m=21$ and
\begin{align*}
g_1 &= \str{2217457753027220341},\\
g_2 &= \str{34136156204421463561},\\
g_3 &= \str{13421655764256421611},\\
g_4 &= \str{53626442356412374141}.
\end{align*}
Let $C_1=Q_6(g_1,\dots,g_4)\subseteq\F_8^{84}$. It is a quasi-cyclic code of
index $4$ with check polynomial
\[
  h=(x+1)(x+w)(x^2+w^3x+w^6)(x^2+w^6x+w^5).
\]
As in Section~\ref{sec:A}, let $G$ be the code obtained by appending to the
word of each $a(x)$ the letter $a(1)=a_0+a_1+\dots+a_5$. A generator matrix
of $G$ consists of the six rows
$(x^ig_1\mid x^ig_2\mid x^ig_3\mid x^ig_4)\bmod(x^{21}-1)$, $i=0,\dots,5$,
each followed by the letter $1$.

Let $C_0\subset C_1$ be the subcode of the words with $a(1)=0$, of dimension
$5$. Enumeration of all words of $C_1$ gives:
\begin{itemize}
\item $C_0$ has minimum distance $68$, with 3\,234 words of weight $68$;
\item the smallest weight of a word of $C_1$ outside $C_0$ is $66$, with
      4\,998 such words.
\end{itemize}
By Lemma~\ref{lem:X} with $d_0=68$, $e=66$ and $s=1$, as in
Section~\ref{sec:A}, $G$ is an $[85,6,67]_8$ code. It has 4\,998 words of
weight $67$.

\begin{remark}\label{rem:G}
The code $C_1$ is a quasi-cyclic $[84,6,66]_8$ code, and $66$ is also the
lower bound in~\cite{codetables} for $n=84$, $k=6$. Let $\beta=h/(x+1)$. The
word of $a=\beta/\beta(1)$ is
$j=(w^{21}\mid (w^2)^{21}\mid (w^5)^{21}\mid (w^4)^{21})$ and has $a(1)=1$. So
$G=\{(c+tj\mid t)\}$ with $j$ constant on each block, as in
Remark~\ref{rem:A}.
\end{remark}

\begin{remark}\label{rem:Gprior}
The code listed in~\cite{codetables} for $n=84$, $k=6$ is also a
quasi-cyclic $[84,6,66]_8$ code of index $4$ with $m=21$; it is credited
there to Hristov, and these parameters are among those
of~\cite{Hristov2008}. It is a different code from $C_1$: it has 5\,145
words of weight $66$, whereas $C_1$ has 4\,998. Up to scalar multiples there
are $(8^6-1)/7=37\,449$ nonzero columns. None of them extends the listed
code to minimum distance $67$, and exactly one extends $C_1$, the all-ones
column used above.

The lower bound $66$ for $n=85$, $k=6$ is credited in~\cite{codetables} to
Maruta, Shinohara and Takenaka~\cite{MST2008}. Their $[85,6,66]_8$ code has
777 words of weight $66$. Their $[93,6,73]_8$ code has no set of eight
coordinates whose deletion leaves minimum distance $67$, and the
$[95,6,75]_8$ code of~\cite{codetables} described in
Remark~\ref{rem:hristov} has no such set of ten coordinates. Both
statements come from exhaustive searches over the sets of coordinates.
They also exclude an $[86,6,68]_8$ code obtained by deleting seven
coordinates of the first code or nine of the second, since a set of deleted
coordinates that left minimum distance $68$ could be enlarged by one
coordinate to a set that leaves $67$. Since no column extends the listed
$[84,6,66]_8$ code to minimum distance $67$, no two appended columns extend
it to minimum distance $68$. We did not rebuild the $[89,6,70]_8$ code from
which~\cite{codetables} obtains its codes for $n=85$ and $n=86$ by deleting
the last four and the last three coordinates, so we do not know what other
sets of deleted coordinates give.
\end{remark}

\subsection{A \texorpdfstring{$[78,8,56]_7$}{[78,8,56]\_7} code}\label{sec:H}

Let $q=7$, $m=38$ and
\begin{align*}
g_1 &= \str{11134206144316552225610253625541},\\
g_2 &= \str{1112436663014034041643631356204426151}.
\end{align*}
Let $C_0=Q_7(g_1,g_2)\subseteq\F_7^{76}$. It is a quasi-cyclic code of
index $2$ with check polynomial
\[
  h=(x+1)(x^3+2x^2+1)(x^3+2x+6),
\]
the product of $x+1$ and two irreducible polynomials of degree $3$. Let
$j=1^{76}$ be the all-ones vector. It is not in $C_0$. Enumeration gives:
\begin{itemize}
\item $C_0$ is a $[76,7,56]_7$ code with 1\,596 words of weight $56$
      (the lower bound in~\cite{codetables} for $n=76$, $k=7$ is also $56$);
\item the smallest weight of a word of $C_0+\langle j\rangle$ outside $C_0$
      is $54$, with 2\,964 such words.
\end{itemize}
By Lemma~\ref{lem:X} with $d_0=56$, $e=54$ and $s=2$, the code
\[
  H=\{\,(c+tj\mid t,t) \;:\; c\in C_0,\ t\in\F_7\,\}
\]
is a $[78,8,56]_7$ code. It has $1\,596+2\,964=4\,560$ words of weight $56$. A
generator matrix consists of the seven rows of $C_0$, each followed by
$0,0$, and the row $(j\mid 1,1)$.

\begin{remark}\label{rem:H}
The code listed in~\cite{codetables} for $n=78$, $k=8$ is a quasi-cyclic
$[80,8,57]_7$ code of index $5$ with $m=16$, with two coordinates deleted.
That $[80,8,57]_7$ code has 2\,208 words of weight $57$, and every three
coordinates are all nonzero in one of these words. So deleting two
coordinates of that code always gives minimum distance $55$, and deleting
three always gives $54$, the value listed for $n=77$. The
quasi-cyclic $[80,8,56]_7$ code of Daskalov and
Gulliver~\cite{DaskalovGulliver2001} has 600 words of weight $56$, and
every two coordinates are both nonzero in one of them, so deleting two of
its coordinates always gives minimum distance $54$.
\end{remark}

\begin{proof}[Proof of Corollary~\ref{cor:89}]
Delete the last coordinate of $B$: by enumeration the resulting code has
minimum distance $64$. Alternatively, add a zero coordinate to $A$. In
Section~\ref{sec:D}, Lemma~\ref{lem:X} with $s=1$ and with $s=0$ gives a
$[96,8,71]_8$ code and the $[95,8,70]_8$ code $C_0+\langle j\rangle$.
Deleting any one coordinate of the latter gives, by enumeration, a code of
minimum distance $69$, hence a $[94,8,69]_8$ code. In
Section~\ref{sec:H}, Lemma~\ref{lem:X} with $s=1$ gives a $[77,8,55]_7$
code, which is $H$ with its last coordinate deleted. In
Section~\ref{sec:G}, Lemma~\ref{lem:X} with $s=2$ gives an $[86,6,68]_8$
code, which is $G$ with its last coordinate repeated; it has
$3\,234+4\,998=8\,232$ words of weight $68$.
\end{proof}

Apart from Corollary~\ref{cor:89}, deleting one coordinate of one of the
eight codes, shortening it at one coordinate, or adding a zero coordinate
gives no further new code. For
each of $A$, $B$, $C$, $D$, $E$, $F$, $G$ and $H$, enumeration shows that
deleting any single coordinate lowers the minimum distance by exactly one.
Shortening at any single coordinate leaves the minimum distance unchanged,
except that shortening $G$ at its last coordinate gives the code $C_0$ of
Section~\ref{sec:G} and minimum distance $68$. Deleting a
coordinate of $B$, of $D$ or of $H$ gives parameters of Corollary~\ref{cor:89}, and
deleting a coordinate of $A$ gives $[87,8,63]_7$, the parameters of
Remark~\ref{rem:canli}. The other resulting parameters are
$[87,7,64]_7$, $[89,7,65]_7$, $[91,8,65]_7$, $[97,6,75]_8$, $[97,7,74]_8$,
$[99,7,75]_8$, $[96,7,72]_8$, $[98,8,72]_8$, $[43,6,31]_9$, $[43,7,30]_9$,
$[45,7,31]_9$, $[96,8,70]_7$, $[96,7,71]_7$, $[98,8,71]_7$, $[84,6,66]_8$,
$[84,5,67]_8$, $[84,5,68]_8$, $[86,6,67]_8$, $[77,7,56]_7$ and
$[79,8,56]_7$; they do not
exceed the lower bounds $65$, $66$, $65$,
$77$, $74$, $75$, $73$, $72$, $32$, $30$, $31$, $70$, $72$, $71$, $66$,
$68$, $68$, $67$, $57$ and $56$ listed
in~\cite{codetables}. Every two coordinates of the $[95,8,70]_8$ code of
Corollary~\ref{cor:89} are both nonzero in some word of weight $70$, so
deleting two of its coordinates gives the parameters $[93,8,68]_8$, which
the table already lists.

\section{Verification and data}\label{sec:verification}

The minimum distance of each of the eight codes $A$ to $H$ was computed in
three ways
that share no code:
\begin{enumerate}
\item with a C implementation of the Brouwer--Zimmermann
      algorithm~\cite{Grassl2006,Betten2006};
\item with the function \texttt{minimum\_distance} of
      SageMath~10.10~\cite{sagemath};
\item by listing all $q^k$ codewords with the Python program
      \texttt{verify.py}, an ancillary file of this paper.
\end{enumerate}
All three give $64$, $65$, $75$, $72$, $31$, $71$, $67$ and $56$. For the codes obtained
from $D$ by deleting its last coordinate, its last two and its last three
coordinates, all three give $71$, $70$ and $69$. The program
\texttt{verify.py} needs only Python and NumPy and runs in several
minutes. It rebuilds every code from
the strings printed above (and the code of Remark~\ref{rem:canli} from its
polynomials in~\cite{chenqt}, the codes of Remark~\ref{rem:hristov} from
the polynomials in~\cite{codetables} and~\cite{Hristov2008}, and the codes
of Remark~\ref{rem:yu} from the matrix in~\cite{YuRepo}, and the codes of
Remarks~\ref{rem:F}, \ref{rem:Gprior} and~\ref{rem:H} from the data
in~\cite{codetables}, \cite{Hristov2009}, \cite{MST2008}
and~\cite{DaskalovGulliver2001}), and recomputes
every count in
Sections~\ref{sec:A}--\ref{sec:H} and
Remarks~\ref{rem:canli}, \ref{rem:hristov}, \ref{rem:yu}, \ref{rem:F},
\ref{rem:G}, \ref{rem:Gprior} and~\ref{rem:H}, the
minimum distances in the proof of Corollary~\ref{cor:89}, the statements
about single coordinates after that proof, the statements
about the check polynomials, and the full weight distributions, which it
writes to the file \texttt{facts.json}.

The generator matrices and the programs of items~1 and~2 are deposited at
Zenodo~\cite{zenodoA,zenodoB,zenodoC,zenodoD,zenodoE,zenodoF,zenodoG,zenodoH}
and mirrored at
\url{https://github.com/cochinescu/linear-code-records}. The Zenodo records
also include input files for Magma~\cite{magma}. They had
not been run when
the records were published; we then ran them unchanged in the Magma online
calculator (version V2.29-10), which gives the same eight minimum distances.
The ancillary files of
this paper contain the generator matrices of the eight codes as
plain text. The codes $A$ to $E$ were reported to the maintainer
of~\cite{codetables} on 10 November 2026.

Before submission we looked for earlier codes with these parameters
in~\cite{codetables} and~\cite{chenqt}, in the recent papers
\cite{AydinLambrinosVandenBerg2019,Akre2023,Dastbasteh2025}, which report
new codes over $\F_7$, in~\cite{Hristov2008,Aydin2019,He2026}, which report
new codes over $\F_8$, and in
the file~\cite{YuRepo}, which contains codes over $\F_8$ and $\F_9$, and
found none; Remarks~\ref{rem:hristov} and~\ref{rem:yu} describe the closest
codes we found. For $F$, $G$ and $H$ we looked in addition
in~\cite{Hristov2009,MST2008,DaskalovGulliver2001};
Remarks~\ref{rem:F}, \ref{rem:Gprior} and~\ref{rem:H} describe the closest
codes we found. The
database of best known linear codes of Magma V2.29-10, queried on
10 and 12 November 2026, has the same lower bounds as~\cite{codetables} for
the fourteen parameter sets of Theorem~\ref{thm:main} and
Corollary~\ref{cor:89}. This
search was not exhaustive.

\section{How the codes were found}\label{sec:search}

The code $C_1$ of Section~\ref{sec:A} was found by a heuristic search among
one-generator quasi-cyclic codes with the check polynomial $h=(x-1)h_7$,
evaluated together with the appended coordinate $a(1)$. Remark~\ref{rem:A} then suggested the general recipe of
Lemma~\ref{lem:X}: take a good quasi-cyclic code $C_0$ whose check polynomial
is not divisible by $x-1$, add a vector that is constant on each block, and
append a repetition code. A scan of such vectors over quasi-cyclic codes that
meet the lower bounds of~\cite{codetables} produced the code $C$ of
Section~\ref{sec:C}. The codes $D$ and $E$ came from the same scan applied
to quasi-cyclic codes with reducible check polynomials, found by the same
heuristic search; as Remark~\ref{rem:D} shows, the check polynomial may be
divisible by $x-1$, as long as the added vector is not in the code. The
codes $F$ and $H$ came from this scan as well, applied to quasi-cyclic
codes $C_0$ found by the same heuristic search. The code $C_1$ of
Section~\ref{sec:G} came from the heuristic search, evaluated together with
the appended coordinate $a(1)$ as in Section~\ref{sec:A}; that
Lemma~\ref{lem:X} with $s=2$ then gives the $[86,6,68]_8$ code of
Corollary~\ref{cor:89} was pointed out by one of the AI systems used to
review a draft of this version. The two
columns of Section~\ref{sec:B} were found by a
search for two coordinates such that every word of minimum weight is nonzero
in at least one of them, as in~\cite{Kohnert2009}. None of these
searches was exhaustive, and we make no claim about what else these families
contain.

\paragraph{Use of AI tools.}
The search and verification programs were written and run by AI coding
agents based on large language models, working under the author's direction
on two personal computers, and this text was drafted with AI assistance. Before
release, each code was also recomputed by two further AI systems, each with
a program written from scratch. Machine learning has been used before in the
search for champion codes~\cite{He2026}. The author has checked the
statements of this paper and takes responsibility for them; they do not
rest on trust in these tools, since each can be confirmed with the ancillary
program.

\paragraph{Acknowledgements.}
This work depends on the tables of M.~Grassl~\cite{codetables} and on the
database of E.~Z.~Chen~\cite{chenqt}.

\end{document}